# Unity gain and non-unity gain quantum teleportation

W. P. Bowen[1], N. Treps[1], B. C. Buchler[1], R. Schnabel[1], T. C. Ralph[2], T. Symul[1] and P. K. Lam[1]

*(Invited Paper)*

*Abstract*—We investigate continuous variable quantum teleportation. We discuss the methods presently used to characterize teleportation in this regime, and propose an extension of the measures proposed by Grangier and Grosshans [1], and Ralph and Lam [2]. This new measure, the gain normalized conditional variance product $\mathcal{M}$, turns out to be highly significant for continuous variable entanglement swapping procedures, which we examine using a necessary and sufficient criterion for entanglement. We elaborate on our recent experimental continuous variable quantum teleportation results [3], demonstrating success over a wide range of teleportation gains . We analyze our results using fidelity; signal transfer, and the conditional variance product; and a measure derived in this paper, the gain normalized conditional variance product.

*Index Terms*—Entanglement, squeezing, quantum information, teleportation.

## I. Introduction

QUANTUM teleportation was first proposed by Bennett *et al.* [4] in the discrete variable regime of single photon polarization states. They showed that, by utilizing entanglement, it was possible to perform a perfect quantum reconstruction of a state from classical destructive measurements. This technique in now of significant relevance to quantum information systems in terms of both communicating [5] and processing [6] quantum information. The first experimental demonstrations of quantum teleportation were performed in 1997 on the polarization state of single photons [7]. Quantum teleportation has now been generalized to many other regimes, and has been demonstrated using liquid NMR ensembles [8], and optical field states [3], [9], [10]. Here we consider teleportation of the quadrature amplitudes of a light field [2], [11], [12].

Since the first demonstration of optical field state teleportation by Furusawa *et al.* [9], there has been considerable discussion about how continuous variable quantum teleportation may be performed using different systems [2], [12], [13], [14], [15]; applied to different input states [16], [17]; generalized to multi-party situations [18]; and comprehensively characterized [19], [20]. Given this intense interest, it is somewhat surprising that further continuous variable teleportation experiments have only been performed very recently. Zhang *et al.* [10] performed a detailed analysis and presented new results from the Furusawa et al. setup with a fidelity improvement from 0.58 to 0.61; and Bowen *et al.* [3] presented results from a new teleportation experiment. This paper attempts to summarize some of the discussion about how best to characterize continuous variable teleportation, dividing the process into two regimes: unity gain, and non-unity gain. We discuss existing methods to characterize continuous variable teleportation within each regime, and introduce a new measure for the non-unity gain regime, the gain normalized conditional variance product. A necessary and sufficient analysis of continuous variable entanglement swapping is given to emphasize the importance of the non-unity gain regime, and to demonstrate the significance of the gain normalized conditional variance product. We elaborate on our experimental demonstration of continuous variable teleportation of the amplitude and phase quadratures of an optical field, first published in [3]. We provide additional experimental results and an analysis of our results using the gain normalized conditional variance product.

The teleportation protocol demonstrated here was, in many ways, similar to the one used by Furusawa *et. al* [9]. It consisted of three parts: measurement (Alice), reconstruction (Bob), and generation and verification (Victor). There were some notable differences, however, in both the methods of input state measurement and of output state verification. In our experiment the input and output states were both analyzed by Victor in the same homodyne detector, and in a location spatially separated from Alice and Bob. This spatial separation is in line with the original concept of quantum teleportation, and is important to ensure that Alice and Bob obtain no information about how Victor encodes the input state. If they do obtain information about the encoding they can use it to artificially improve the quality of Bob's reconstructed state. Our experiment is based on a Nd:YAG laser that produces two squeezed beams in two independently pumped optical parametric amplifiers (OPAs). Using independent OPAs reduces the degradation of squeezing caused by green-induced-infrared-absorption which is presently one of the limiting factors in the Furusawa *et al.* teleportation setup [21]. We use a more compact configuration for Alice's measurements which relies on only two detectors and one electronic locking loop, as opposed to the four detectors, two local oscillators, and two locking loops used in the Furusawa *et al.* experiment. The use of two independent modulators each for encoding of Victor's input state and Bob's reconstructed output state allows the phase space of the input state to be completely spanned, and the amplitude and phase quadrature teleportation gains to be accurately experimentally verified. The fidelity especially, is extremely sensitive to gain, so that in our experiment *a posteriori* verification of the applied gain was essential to confirm any fidelity results.

We analyzed the efficacy of our experiment using three measures: fidelity $\mathcal{F}$; a T-V diagram of the signal transfer $T_q$ and conditional variance product $V_q$ between the input and

1 Department of Physics, Faculty of Science, Australian National University, ACT 0200, Australia
2 Department of Physics, Centre for Quantum Computer Technology, University of Queensland, St Lucia, QLD, 4072 Australia

output states; and the gain normalized conditional variance product $\mathcal{M}$. In the unity gain teleportation regime we used the fidelity, $\mathcal{F}$, between the input and output states to characterize the teleportation protocol, and observed an optimum of $\mathcal{F} = 0.64 \pm 0.02$ where $\mathcal{F} \leq 0.5$ when only classical resources are used. The fidelity degrades quickly as the teleportation gain moves away from unity, however, and is not an appropriate measure of non-unity gain teleportation. Instead, we use the signal transfer $T_q$ and conditional vairnace $V_q$ between the input and output states, in a manner analogous to QND analysis [2]. This enables a more detailed two dimensional characterization of the performance of our teleporter. $T_q$ and $V_q$ both have physical significance. $T_q > 1$ ensured that Bob's output state contains more infomation about the input state than any other possible reconstruction, and $T_q = 1$ therefore defines an 'information cloning' limit. $V_q < 1$ is necessary to enable Bob to reconstruct non-classical features of the input state such as squeezing. We observe an optimum signal transfer of $T_q = 1.06 \pm 0.02 > 1$; and simultaneously observe $V_q = 0.96 \pm 0.10 < 1$ and $T_q = 1.04 \pm 0.03 > 1$, at unity teleportation gain this would imply a fidelity surpassing the no-cloning limit. We analyze the gain normalized conditional variance product $\mathcal{M}$ introduced in this paper and demonstrate $\mathcal{M} < 1$ for a gain bandwidth from $g_{\min} = 0.58$ to $g_{\max} = 1.21$, with an optimum of $\mathcal{M} = 0.22 \pm 0.2$. Our teleportation protocol could be used to successfully perform entanglement swapping throughout this bandwidth.

## II. CONTINUOUS VARIABLE TELEPORTATION PROTOCOL

Quantum teleportation is usually described as the disembodied transportation of a quantum state from one place (Alice) to another (Bob). Or in other words, it is a process that allows Bob to reconstruct a quantum state from measurements performed by Alice at time $t_0$, using only classical communication and local operations after time $t_0$ (see fig. 1). The concept of quantum teleportation was first proposed [4] and demonstrated [7] in the discrete variable regime of single, or few, photons. It has since then been generalized to the continuous variable regime [2], [11], [12], which is the setting for this paper. In our work, as in that of ref. [9], the teleported states are modulation sidebands of a bright optical beam. These modulation sidebands can be described using the field annihilation $\hat{a}$ and creation $\hat{a}^\dagger$ operators, where $[\hat{a}, \hat{a}^\dagger] = 1$. The annihilation and creation operators can be expressed in terms of measurable Hermitian operators: $\hat{a} = (\hat{X}^+ + i\hat{X}^-)/2$ and $\hat{a}^\dagger = (\hat{X}^+ - i\hat{X}^-)/2$. Here $\hat{X}^\pm = 2\alpha^\pm + \delta\hat{X}^\pm$ are the amplitude (+) and phase (-) quadrature operators of the field, the coherent amplitude of the field is given by $\alpha = \sqrt{\alpha^{+2} + \alpha^{-2}}$, where $\alpha^\pm = \langle\hat{X}^\pm\rangle/2$ are its real (+) and imaginary (-) parts. $\delta\hat{X}^\pm$ are the phase and amplitude quadrature noise operators and have the commutation relation

$$[\delta\hat{X}^+, \delta\hat{X}^-] = 2i \qquad (1)$$

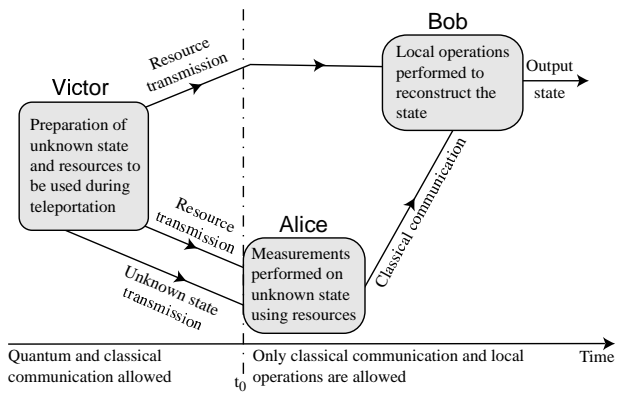

Fig. 1. Time line for a quantum teleportation experiment.

Throughout this paper the variances of these noise operators are denoted by $\Delta^2 \hat{X}^\pm = \langle(\delta\hat{X}^\pm)^2\rangle$. The commutation relation of eq. (1) dictates that $\Delta^2 \hat{X}^+ \Delta^2 \hat{X}^- \geq 1$. This uncertainty product forbids the simultaneous exact knowledge of the amplitude and phase of an optical field and thus prevents perfect duplication or *quantum cloning*.

Until the proposal of Bennett *et. al* [4], uncertainty products were thought to fundamentally limit the quality of any teleportation protocol. Bennett proved that in the discrete regime this was not the case, by utilizing shared entanglement between Alice and Bob. Since then many other teleportation protocols have been proposed [22], all of which rely on shared entanglement. In continuous variable quadrature teleportation protocols [2], [11], [12], Alice and Bob share a quadrature entangled pair, which in this work we restrict to be Gaussian. We generate Gaussian quadrature entanglement by combining two equally amplitude squeezed beams with a $\pi/2$ phase shift on a 50/50 beam splitter[23]. The outputs of this beam splitter are quadrature entangled, with the entanglement evidenced through strong amplitude/amplitude and phase/phase quadrature correlations between the beams. One of these entangled beams is sent to Alice and the other to Bob, as shown in fig. 2. Victor provides an unknown input state to Alice who then interferes it with her entangled beam and performs an amplitude quadrature measurement on one of the outputs and a phase quadrature measurement on the other. The measurement results are sent through classical communication to Bob, who encodes them using amplitude and phase modulators on his entangled beam. The phase and amplitude quadratures of the output state produced can be written

$$\hat{X}^+_{\text{out}} = g^+ \hat{X}^+_{\text{in}} + \frac{(1+g^+)}{\sqrt{2}} \hat{X}^+_{\text{sqz},1} + \frac{(1-g^+)}{\sqrt{2}} \hat{X}^-_{\text{sqz},2} \quad (2)$$

$$\hat{X}^-_{\text{out}} = g^- \hat{X}^-_{\text{in}} + \frac{(1+g^-)}{\sqrt{2}} \hat{X}^+_{\text{sqz},2} + \frac{(1-g^-)}{\sqrt{2}} \hat{X}^-_{\text{sqz},1} \quad (3)$$

where the sub-scripts in and out label the input and output states respectively, $g^\pm = \alpha^\pm_{\text{out}}/\alpha^\pm_{\text{in}}$ are the amplitude and phase quadrature teleportation gains, and the sub-scripts sqz,1 and sqz,2 label the squeezed beams used to generate our quadrature entanglement. Assuming that the entanglement used in the teleportation protocol is produced by two equally amplitude squeezed beams with $\Delta^2 \hat{X}_{\text{sqz}} = \Delta^2 \hat{X}^+_{\text{sqz},1} = \Delta^2 \hat{X}^+_{\text{sqz},2} < 1$ and $\Delta^2 \hat{X}_{\text{anti}} = \Delta^2 \hat{X}^-_{\text{sqz},1} = \Delta^2 \hat{X}^-_{\text{sqz},2} > 1$, the quadrature variances of the output state are

$$\Delta^2 \hat{X}^\pm_{\text{out}} = g^{\pm 2} \Delta^2 \hat{X}^\pm_{\text{in}} + \frac{(1+g^\pm)^2}{2} \Delta^2 \hat{X}_{\text{sqz}} + \frac{(1-g^\pm)^2}{2} \Delta^2 \hat{X}_{\text{anti}} \quad (4)$$

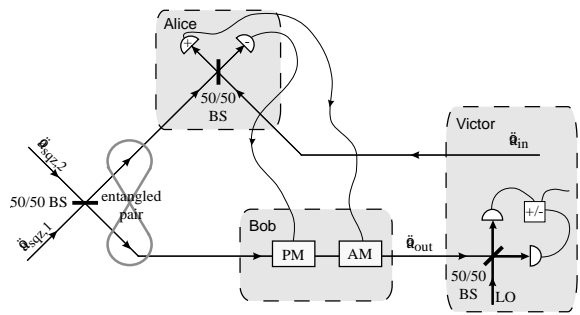

Fig. 2. Continuous variable quantum teleportation protocol, AM: amplitude modulator, PM: phase modulator, BS: beam splitter.

We then see that for unity gain teleportation ($g^\pm = 1$), as $\Delta^2 \hat{X}_{\text{sqz}} \to 0$, the amplitude and phase quadratures of the output state approach those of the input ($\hat{X}^\pm_{\text{out}} \to \hat{X}^\pm_{\text{in}}$). So that, at least in the limit of perfect squeezing, perfect teleportation of the amplitude and phase quadratures of an optical field can be achieved. Of course, unity gain is not necessarily the only interesting operation point of a teleporter. In the following section we discuss methods to characterize the success of teleportation in both the unity gain regime, and in other regimes.

### III. CHARACTERIZATION OF TELEPORTATION

There is some debate in the quantum optics community about specifically what constitutes continuous variable quantum teleportation [1], [2], [19], [20], [24], [25]. One approach is to restrict quantum teleportation to systems that produce output states identical to their input states but with noise convolution, i.e. for all operators involved $\hat{X}_{\text{out}} = \hat{X}_{\text{in}} + \hat{X}_{\text{noise}}$ [20], [24]. We term this *unity gain* teleportation. The interesting aspect of quantum teleportation is that classical communication can be used to transmit quantum information. This can also be demonstrated in situations where the unity gain condition is not true. Indeed under certain conditions optimum transmission of quantum information occurs for non-unity gain. We term this class of teleportation protocols *non-unity gain* teleportation.

A number of methods have been proposed to measure the success of quantum teleportation in both unity gain [20], [26], and non-unity gain [2], [25] situations. These measures typically adopt either a *system dependent* or *state dependent* approach. System dependent measures such as those introduced by Grangier and Grosshans [1] and Ralph and Lam [2] characterize the effect of the system on the input state, such as what noise it introduces, in a manner independent of the form of the input state. While state dependent measures such as the fidelity [26], quantum interferometry [25], and entanglement swapping [16] formulate relationships between the input and output states that cannot be satisfied through any classical means, and vary according to the form of the input state as well as the quality of teleportation. The transfer function approach of system dependent measures is perhaps more useful for characterization of quantum communication networks; and the state dependent approach more relevant when fragile quantum states are being teleported. Ultimately, however, these two approaches should yield equivalent results.

*A. Fidelity*

The most well known and widely used measure of the success of a teleportation protocol is the *fidelity* of teleportation [26], which is state dependent. Fidelity measures the state-overlap between the input $|\psi_{\text{in}}\rangle$ and output $\hat{\rho}_{\text{out}}$ states, and is given by $\mathcal{F} = \langle\psi_{\text{in}}|\hat{\rho}_{\text{out}}|\psi_{\text{in}}\rangle$. $\mathcal{F} = 1$ indicates that the output state is a perfect reconstruction of the input, and $\mathcal{F} = 0$ if the input and output states are orthogonal. The maximum fidelity achievable classically lies somewhere between these bounds and depends strongly on the input state. If the input state is coherent, and the entanglement resource and all other noise sources are Gaussian, the fidelity is given by

$$\mathcal{F} = \frac{2e^{-k^+ - k^-}}{\sqrt{(1+\Delta^2 \hat{X}^+_{\text{out}})(1+\Delta^2 \hat{X}^-_{\text{out}})}} \quad (5)$$

where $k^\pm = \alpha^{\pm 2}_{\text{in}}(1-g^\pm)^2/(1+\Delta^2 \hat{X}^\pm_{\text{out}})$. If the input state is selected from a sufficiently broad Gaussian distribution of coherent states (i.e, a set having a range coherent amplitudes $\Delta \alpha^\pm_{\text{in}} \gg 1$) in a manner unknown to Bob, the generally accepted classical limit to fidelity is $\mathcal{F}_{\text{class}} = 1/2$. This classical limit is determined by considering the optimal teleportation protocol without the availability of an entanglement resource when a weighted average of the fidelity over the input state distribution is made. There is some debate however, about whether the *no-cloning limit* is more appropriate [20] since in the original teleportation paper of Bennett *et. al* they state "Of course Alices original $|\phi\rangle$ [$|\psi_{\text{in}}\rangle$ here] is destroyed in the process, as it must be to obey the no-cloning theorem" [4]. For a sufficiently broad distribution of coherent states the no-cloning limit occurs at $\mathcal{F}_{\text{no-cloning}} = 2/3$. Achieving $\mathcal{F} = 2/3$ ensures that Bob's reconstruction of the input state is better than any other possible reconstruction. There is, therefore, significance in experimentally surpassing both $\mathcal{F} = 1/2$ and $\mathcal{F} = 2/3$.

Notice that the fidelity of eq. (5) has an exponential dependence on $k^+$ and $k^-$. The condition $\Delta\alpha^\pm_{\text{in}} \gg 1$ (and therefore for some input states $\alpha^\pm_{\text{in}} \gg 1$) for validity of $\mathcal{F}_{\text{class}}$ and $\mathcal{F}_{\text{no-cloning}}$, then results in a very strong dependence of the fidelity on gain. The optimum fidelity occurs at $g^\pm \approx 1$, and for the ideal case of an infinitely broad set of input states ($\Delta\alpha^\pm_{\text{in}} \to \infty$), at $g^\pm = 1$. In a physically realistic situation however, there will be some error associated with $g^\pm$. The effect of this error on the fidelity as a function of the input coherent amplitude is shown in fig. 3. No matter how small the gain error, as $\alpha_{\text{in}}$ increases the fidelity falls away towards zero. This puts an upper limit on the breadth of the distribution of states that the teleporter can handle and causes the values of $\mathcal{F}_{\text{class}}$ and $\mathcal{F}_{\text{no-cloning}}$ to become somewhat higher than the limits given above [27], [28]. In experiments to date, these issues have been avoided by strictly defining the fidelity as valid only when $g^{\pm} = 1$. In that case $k^{\pm} = 0$, and the fidelity becomes independent of $\alpha_{\rm in}$. Given this definition of fidelity, it is critical to experimentally verify $g^{\pm}$.

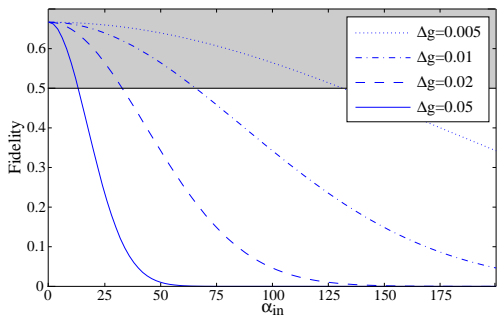

Fig. 3. Fidelity as a function of coherent amplitude $\alpha_{\rm in}$ for a range of teleportation gain errors $\Delta g = g - 1$, with 3 dB of squeezing ($\Delta^2 \hat{X}_{\rm sqz} = 0.5$).

### B. The conditional variance product and signal transfer

From a system dependent perspective, the ideal way to characterize a teleportation protocol is to identify exactly the transfer function of the protocol between the input and output states. This approach is relatively easy for protocols utilizing Gaussian entanglement, which is the only form of continuous variable entanglement presently experimentally available. For the continuous variable quadrature teleportation discussed in this paper, only two variables for each quadrature need to be characterized to completely define the transfer function of the system. They are the teleportation gain, and the amount of noise (or degradation) introduced during the teleportation process. All that remains to be done is then to define bounds on the system that are impossible to exceed classically.

The first such approach was the work of Ralph and Lam [2]. They proposed a characterization in terms of the conditional variance between the input and output states $\Delta^2 \hat{X}_{\rm in|out}^{\pm} = \Delta^2 \hat{X}_{\rm out}^{\pm} - |\langle \delta \hat{X}_{\rm in}^{\pm} \delta \hat{X}_{\rm out}^{\pm} \rangle|^2 / \Delta^2 \hat{X}_{\rm in}^{\pm}$, and the signal transfer from the input to the output state $T^{\pm} = {\rm SNR}_{\rm out}^{\pm} / {\rm SNR}_{\rm in}^{\pm}$ where ${\rm SNR}^{\pm} = \alpha^{\pm} / \Delta^2 \hat{X}^{\pm}$ are conventional signal-to-noise ratios. The conditional variance measures the noise introduced during the protocol, and the signal transfer is related to the gain of teleportation.

Ralph and Lam demonstrated limits to both the joint signal transfer, and the joint conditional variance of the amplitude and phase quadratures that, in a teleportation protocol, can only be overcome by utilizing entanglement. Alice's measurement is limited by the generalized uncertainty principle $\Delta^2 \hat{M}^+ \Delta^2 \hat{M}^- \geq 1$ [29], where $\Delta^2 \hat{M}^{\pm}$ are the measurement penalties which holds for simultaneous measurements of non-commuting quadrature amplitudes. In the absence of entanglement this places a strict limit on Bob's reconstruction accuracy which, in terms of quadrature signal transfer coefficients $T^{\pm}$, can be expressed as

$$T_q = T^+ + T^- - T^+ T^- \left(1 - \frac{1}{\Delta^2 \hat{X}_{\rm in}^+ \Delta^2 \hat{X}_{\rm in}^-}\right) \leq 1 \quad (6)$$

For minimum uncertainty input states ($\Delta^2 \hat{X}_{\rm in}^+ \Delta^2 \hat{X}_{\rm in}^- = 1$), this expression reduces to $T_q = T^+ + T^-$. Perfect signal transfer would give $T_q = 2$.

Bob's reconstruction must be carried out on an optical field, the fluctuations of which obey the uncertainty principle. In the absence of entanglement, these intrinsic fluctuations remain present on any reconstructed field. Therefore, since the amplitude and phase conditional variances $\Delta^2 \hat{X}_{\rm in|out}^{\pm}$ measure the noise added during the teleportation process, they must satisfy $\Delta^2 \hat{X}_{\rm in|out}^+ \Delta^2 \hat{X}_{\rm in|out}^- \geq 1$. This can be written in terms of the quadrature variances of the input and output states and the teleportation gain as

$$V_q = \left(\Delta^2 \hat{X}_{\rm out}^+ - g^{+^2} \Delta^2 \hat{X}_{\rm in}^+\right) \left(\Delta^2 \hat{X}_{\rm out}^- - g^{-^2} \Delta^2 \hat{X}_{\rm in}^-\right) \geq 1 \quad (7)$$

A teleportation protocol that introduced no reconstruction noise would give $V_q = 0$. In the original paper of Ralph and Lam [2] they propose $\Delta^2 \hat{X}_{\rm in|out}^+ + \Delta^2 \hat{X}_{\rm in|out}^- \geq 2$ as the conditional variance limit. For cases where both quadratures are symmetric, such as those considered by them [2], [16], both limits are equivalent. The product limit, however, is significantly more immune to asymmetry in the teleportation gain and provides a more rigorous bound in situations when the entanglement used in the protocol is asymmetric, we therefore prefer it here.

The criteria of eqs. (6) and (7) enable teleportation results to be represented on a T-V graph similar to those used to characterize quantum non-demolition experiments [30]. The T-V graph is two dimensional, and therefore conveys more information about the teleportation process than single dimensional measures such as fidelity. It tracks the quantum correlation and signal transfer in non-unity gain situations. It identifies two particularly interesting regimes that are not evident from a fidelity analysis; the situation where the output state has minimum additional noise ($\min_g \{V_q\}$) which occurs in the regime of $g \leq 1$, and the situation when the input signals are transferred to the output state optimally ($\max_g \{T_q\}$) which occurs in the regime of $g \geq 1$.

Both the $T_q$ and $V_q$ limits have independent physical significance. If a signal has some inherent signal-to-noise ratio, and the noise is truly an unknown quantity, then even in a classical world, that signal-to-noise ratio can in no way be *a posteriori* enhanced. In the case of the teleportation protocol discussed here, Bob receives two signals, the amplitude and phase quadratures of a light field, with a total possible signal transfer of $T_{q,\max} = 2$. If Bob receives $T_{q,{\rm Bob}} = \xi$, then the most signal any other party can receive is $T_{q,{\rm other}} = 2 - \xi$. Therefore, if Bob surpasses $T_q = 1$ then he has received over half of the signal from Alice, and this forbids any others parties from doing so. This is an 'information cloning' limit that is particularly relevant in light of recent proposals for quantum cryptography [31]. Furthermore, if Bob passes the $T_q$ limit at unity gain ($g^{\pm} = 1$), then Bob has beaten the no-cloning limit and has $\mathcal{F} \geq 2/3$. Surpassing the $V_q$ limit is a necessary pre-requisite for reconstruction of non-classical features of the input state such as squeezing. The T-V measure coincides with the teleportation no-cloning limit when both $T_q = V_q = 1$. Clearly it is desirable that the $T_q$ and $V_q$ limits are simultaneously exceeded.

*1) Information cloning and eavesdropper attacks:* Consider that an eavesdropper (Eve) performs an attack on Bob's entangled beam utilizing a beam splitter tap off, whilst allowing Alice to make full used of her entangled beam. As discussed above, Bob can guarantee that the signal transfer to his output state is better than to Eve's if he finds $T_q \geq 1$. Alice then, obliviously, performs her measurements and transmits the results to Bob - a transmission intercepted by Eve. Bob and Eve both then attempt to reconstruct the input state from Alice's measurements and their part of the entanglement. Fig. 4 a) shows Bob's T-V analysis of the teleportation protocol for various entanglement strengths with Eve tapping off 50 % of his entanglement. In this special case, we find that $T_q \leq 1$ for all entanglement strengths ($\Delta^2 \hat{X}_{\text{sqz}}$) and all teleportation gains $g$, with the equality $T_q = 1$ achieveable for any entanglement strength at some gain. In fact, Bob is able to map out the entire physically realistic region of the T-V diagram with $T_q \leq 1$. Since the arrangement is symmetric, Eve obtains the same result. This special case defines the transition point between Bob succesfully surpassing the information cloning limit, and Eve surpassing it.

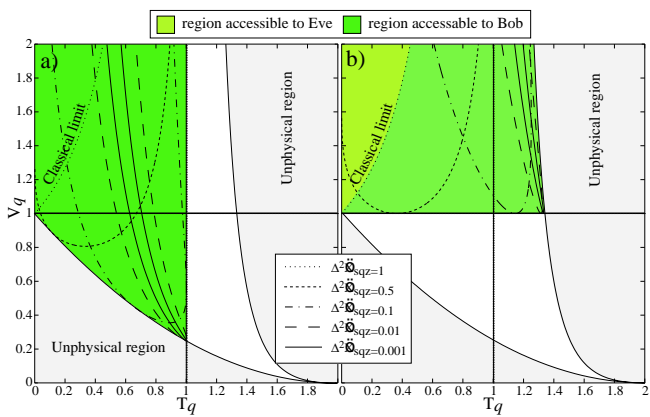

Fig. 4. T-V diagram representation of teleportation with 50% loss introduced to Bob's entangled beam. The lines show the results of the protocol utilizing various squeezing strengths as a function of teleportation gain. If the loss originates from a tap-off introduced by Eve, she at most achieve the same maximum information transfer of $T_q \leq 1$ as Bob.

It is interesting to consider the complimentary situation where Eve taps off a part of Alice's entanglement. In this case Eve can, again, use Alice's measurement results to attempt to reconstruct the input state. This time, however, the situation is not symmetric for Bob and Eve, and we find that Bob can surpass the information cloning limit. The conditional variance between his output state and the input is, however, limited to $V_q \geq 1$. For this situation we see from Fig. 4 b) that Bob can obtain results throughout the physically realistic region of the T-V diagram with $V_q \geq 1$. If Eve attempts to reconstruct the input state, however, she is restricted to the classical region of the T-V diagram.

### C. A gain normalized conditional variance product

In their paper [1], Grangier and Grosshans discuss in some detail both the fidelity, and signal transfer and conditional variance, measures for teleportation. They restrict teleportation to occur only in the unity gain regime. In this case, the signal transfer and conditional variance criteria of eqs. (6) and (7) become equivalent and Grosshans and Grangier propose that both are good measures of teleportation. One advantage that both their measure and fidelity have over the T-V diagram is that they provide a single number for the quality of the teleportation protocol. Although this does provide less information about the operation of the protocol, it allows different teleportation schemes to be directly compared. It is interesting to consider whether a single number can be used to characterize teleportation in the non-unity gain regime. In the following section we will briefly reproduce Grosshans and Grangier's main results and generalize them to the non-unity gain regime.

Grosshans and Grangier quite generally denote the joint measurements $M^\pm$ performed by Alice on the input state as $M^\pm = \Lambda^\pm \hat{X}_{\text{in}}^\pm + \hat{N}_{\text{Alice}}^\pm$ where $\Lambda^\pm$ are amplitude and phase quadrature gains related to the detection, and $\hat{N}_{\text{Alice}}^\pm$ is the noise introduced during the measurement. Both $M^\pm$ are detected photocurrents so that $[M^+, M^-] = 0$. Given that the measurement noise is uncorrelated to the input state, one can easily obtain the Heisenberg uncertainty product

$$\Delta^2 \hat{N}_{\text{Alice}}^+ \Delta^2 \hat{N}_{\text{Alice}}^- \geq |\Lambda^+ \Lambda^-|^2 \qquad (8)$$

and the signal transfer co-efficient from the input state to Alice's photo-currents are

$$T_{\text{Alice}}^\pm = \frac{\Lambda^{\pm\,2} \Delta^2 \hat{X}_{\text{in}}^\pm}{\Lambda^{\pm\,2} \Delta^2 \hat{X}_{\text{in}}^\pm + \Delta^2 \hat{N}_{\text{Alice}}^\pm} \qquad (9)$$

Notice that $T^\pm \leq 1$ always, so that as stated earlier, $T_q$ is bounded from above by two. Using the Heisenberg uncertainty product of eq. (8) it is then possible to derive the same signal transfer bound as given in eq. (6). Of course, Bob still has to reconstruct the input state on his output optical field. The output quadrature operators can be expressed as

$$\hat{X}_{\text{out}}^\pm = \Upsilon^\pm (\Lambda^\pm \hat{X}_{\text{in}}^\pm + \hat{N}_{\text{Alice}}^\pm) + \hat{N}_{\text{Bob}}^\pm \qquad (10)$$
$$= g^\pm \hat{X}_{\text{in}}^\pm + \hat{N}_{\text{total}}^\pm \qquad (11)$$

where $\hat{N}_{\text{Bob}}^\pm$ are quadrature operators describing Bob's initial optical field, $\Upsilon^\pm$ are the amplitude and phase feed-forward gains applied to Alice's measurements, and the teleportation gains $g^\pm = \Upsilon^\pm \Lambda^\pm$. $\hat{N}_{\text{total}}^\pm = \Upsilon^\pm \hat{N}_{\text{Alice}}^\pm + \hat{N}_{\text{Bob}}^\pm$ describe the total noise added to the amplitude and phase quadratures during the teleportation process, and for each quadrature the total noise variance is equal to the condition variance between the input and output states, $\Delta^2 \hat{N}_{\text{total}}^\pm = \Delta^2 \hat{X}_{\text{in}|\text{out}}^\pm$. Grosshans and Grangier observe that this output form dictates the Heisenberg uncertainty product

$$V_q = \Delta^2 \hat{X}_{\text{in}|\text{out}}^+ \Delta^2 \hat{X}_{\text{in}|\text{out}}^- \geq (g^+ g^- - 1)^2 \qquad (12)$$

This implies that unless $g^+ g^- = 1$ the output state must be degraded by some noise. They then restrict their analysis to the unity gain case ($g^+ = g^- = 1$) and, like Ralph and Lam [2], derive $V_q < 1$ as a classical limit for unity gain teleportation.

In this section we wish to extend Grosshans and Grangier's analysis to provide a measure of non-unity gain teleportation based on $V_q$. Since $[\hat{N}_{\text{Bob}}^+, \hat{N}_{\text{Bob}}^-] = 2i$, $\Delta^2 \hat{N}_{\text{Bob}}^+ \Delta^2 \hat{N}_{\text{Bob}}^- \geq 1$.

In general, the noise introduced by Alice and Bob can be broken down into a part due to classical sources such as for example electrical pick-up (sub-script $c$), and a part due to quantum fluctuations (sub-script $q$) that are uncorrelated with each other, $\hat{N}_{\text{Bob}}^{\pm} = \hat{N}_{\text{Bob},c}^{\pm} + \hat{N}_{\text{Bob},q}^{\pm}$ and $\hat{N}_{\text{Alice}}^{\pm} = \hat{N}_{\text{Alice},c}^{\pm} + \hat{N}_{\text{Alice},q}^{\pm}$. Since we are interested in the lower bounds of $V_q$ achievable without entanglement we neglect the classical noise term here, setting $\hat{N}_{\text{Bob}}^{\pm} = \hat{N}_{\text{Bob},q}^{\pm}$ and $\hat{N}_{\text{Alice}}^{\pm} = \hat{N}_{\text{Alice},q}^{\pm}$. In that case, if the noise introduced by Alice and Bob is separable (i.e. not entangled)

$$\begin{aligned} V_q &= \left(\Upsilon^{+\,2}\Delta^2\hat{N}_{\text{Alice}}^+ + \Delta^2\hat{N}_{\text{Bob}}^+\right)\left(\Upsilon^{-\,2}\Delta^2\hat{N}_{\text{Alice}}^- + \Delta^2\hat{N}_{\text{Bob}}^-\right) \\ &\geq |g^+ g^-| + 1 + \\ &\quad \Upsilon^{+\,2}\Delta^2\hat{N}_{\text{Alice}}^+ \Delta^2\hat{N}_{\text{Bob}}^- + \frac{(\Lambda^+\Lambda^-\Upsilon^-)^2}{\Delta^2\hat{N}_{\text{Alice}}^+ + \Delta^2\hat{N}_{\text{Bob}}^-} \end{aligned} \quad (13)$$

Some simple calculus shows that the right-hand-side of inequality (13) is minimized when $\Delta^2\hat{N}_{\text{Alice}}^+ \Delta^2\hat{N}_{\text{Bob}}^- = |\Lambda^+\Lambda^-\Upsilon^-/\Upsilon^+|$, and we find that

$$V_q \geq \left(|g^+ g^-| + 1\right)^2 \quad (14)$$

For $g^+ = g^- = 0$ this classical limit is equivalent to that given in eq. (7), and for non-zero gain it is stronger. We define the non-unity gain teleportation measure $\mathcal{M}$

$$\mathcal{M} = \frac{V_q}{\left(|g^+ g^-| + 1\right)^2} \quad (15)$$

where $\mathcal{M} < 1$ can only be achieved if the noise introduced by Alice and Bob is entangled. It is relatively easy to show that $\mathcal{M} = 1$ is achievable in the teleportation protocol discussed here, using no shared entanglement and for any teleportation gain. For non-unity gain ($g^+ \neq 1$ and/or $g^- \neq 1$) however, Alice and Bob must both utilize local squeezing resources in their measurement and reconstruction processes.

Eq. (12) defines the minimum amount of noise added to the reconstructed state as a function of teleportation gain. This relationship dictates minimum physically achievable values for $\mathcal{M}$ as a function of gain.

$$\mathcal{M}_{\min} = \frac{(g^+ g^- - 1)^2}{(|g^+ g^-| + 1)^2} \quad (16)$$

Fig. 5 shows $\mathcal{M}$ as a function of gain for the teleportation protocol discussed in section II, with a range of utilized squeezing strengths. The optimum of $\mathcal{M}$ always occurs at $g = \sqrt{g^+ g^-} = 1$, and improves as the strength of the entanglement used in the protocol increases (as $\Delta^2\hat{X}_{\text{sqz}} \to 0$). Notice that $\mathcal{M} = 0$ is only possible at $g = 1$, so that at $g \neq 1$ the output state, no matter what entanglement strength is utilized, will incur some reconstruction noise. It is interesting, however, that for all entanglement strengths $\mathcal{M}$ is less than unity for a wide range of gains. Some algebra shows that the gain extremema $g_{\min}$ and $g_{\max}$ for which $\mathcal{M} < 1$ are given by

$$g_{\max/\min} = \frac{\Delta^2\hat{X}_{\text{anti}} - \Delta^2\hat{X}_{\text{sqz}}}{\Delta^2\hat{X}_{\text{anti}} + \Delta^2\hat{X}_{\text{sqz}} - 2} \\ \pm \sqrt{\left(\frac{\Delta^2\hat{X}_{\text{anti}} - \Delta^2\hat{X}_{\text{sqz}}}{\Delta^2\hat{X}_{\text{anti}} + \Delta^2\hat{X}_{\text{sqz}} - 2}\right)^2 - 1} \quad (17)$$

Perhaps contrary to intuition, as the entanglement strength increases (as $\Delta^2\hat{X}_{\text{sqz}} \to 0$) the range of gain over which $\mathcal{M} < 1$ is satisfied decreases. In fact, if a teleportation protocol is intended to operate with a specific gain, a generally non-ideal entanglement strength ($\Delta^2\hat{X}_{\text{sqz,opt}} \neq 0$) exists for maximum efficacy of the protocol. If $g = 1$ then $\Delta^2\hat{X}_{\text{sqz,opt}} = 0$, but as $g \to 0$ or $g \to \infty$, $\Delta^2\hat{X}_{\text{sqz,opt}} \to 1$.

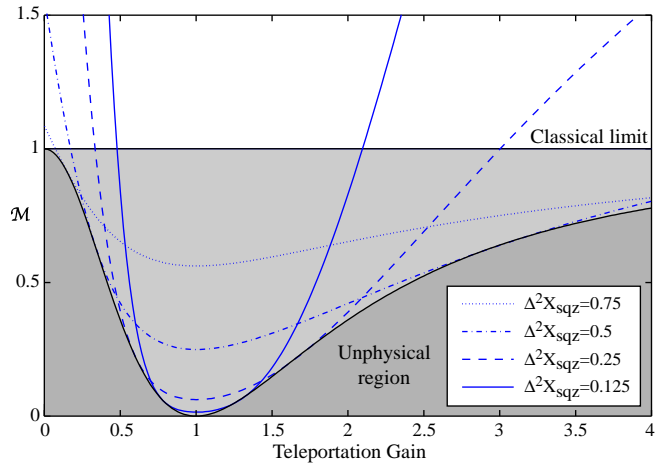

Fig. 5. $\mathcal{M}$ as a function of teleportation gain for pure teleporter input squeezing ($\Delta^2\hat{X}_{\text{sqz}}\Delta^2\hat{X}_{\text{anti}} = 1$) with a range of strengths. The unphysical region here is defined by eq. (16).

### D. A comparison of fidelity, the T-V diagram, and the gain normalized conditional variance product

It is interesting to compare T-V based measures of teleportation with state dependent measures such as fidelity. The gain normalized conditional variance product $\mathcal{M}$ introduced in the previous section can be directly compared to fidelity. It can be seen from eq. (15) that $\mathcal{M}$, and in particular its gain bandwidth, is independent of the coherent amplitude $\alpha$ of the input state. This is not the case for fidelity, and as discussed earlier, it is this dependence that restricts the fidelity to $g = 1$. Fig. 6 shows the fidelity of teleportation for $\Delta^2\hat{X}_{\text{sqz}} = 0.125$ and a range of coherent amplitudes. The gain bandwidth of fidelity clearly depends very strongly on the input coherent amplitude, and in the limit of $\alpha \to \infty$ the gain bandwidth approaches zero, centered around $g = 1$.

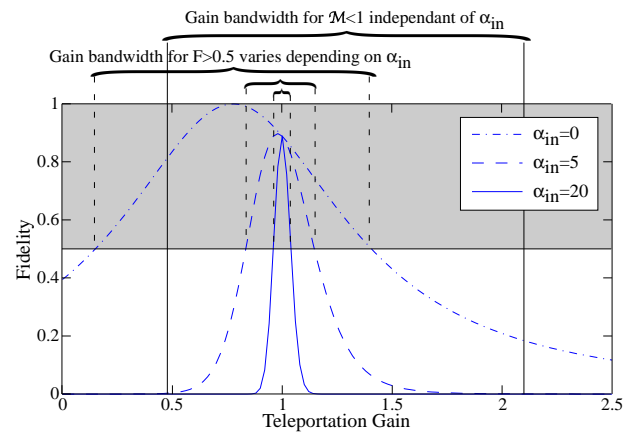

Fig. 6. Fidelity as a function of teleportation gain for $\Delta^2\hat{X}_{\text{sqz}} = 0.125$, $\Delta^2\hat{X}_{\text{anti}} = \Delta^2\hat{X}_{\text{sqz}}^{-1}$, and a range of input coherent amplitudes $\alpha_{\text{in}}$.

Fig. 7 shows a similar result obtained through plotting fidelity contours directly on the T-V diagram. Here we show the fidelity contours at $\mathcal{F} = 1/2$ and $\mathcal{F} = 2/3$ for a range of input coherent amplitudes. From fig. 7 a) we see that with no coherent amplitude the fidelity can be greater than $1/2$, or even $2/3$, for a large area of the T-V diagram, even in the purely classical region in the top left corner of the diagram. As the input coherent amplitude increases through fig. 7 b), c) and d) the area of the T-V diagram in which $\mathcal{F} > 1/2$ or $\mathcal{F} > 2/3$ collapses down to the line defining $g = 1$. We see that for large input coherent amplitudes $\mathcal{F} > 1/2$ can not be achieved in the classical region of the T-V diagram, and $\mathcal{F} > 2/3$ can only be achieved if both $V_q < 1$ and $T_q > 1$. Of course, if we restrict ourselves to the unity gain line then this is true for all coherent amplitudes.

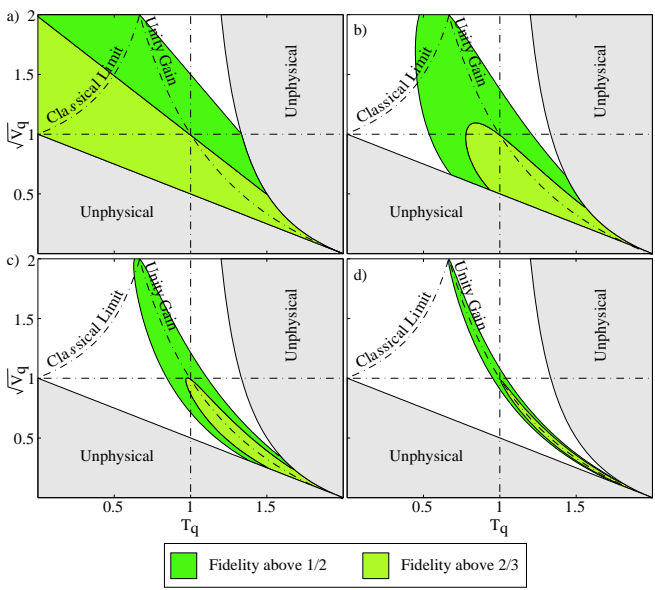

Fig. 7. Fidelity contours for $\mathcal{F} = 1/2$ and $\mathcal{F} = 2/3$ and several conherent amplitudes, represented on the T-V diagram. a) $\alpha_{\text{in}} = 0$, b) $\alpha_{\text{in}} = 2$, c) $\alpha_{\text{in}} = 5$, and d) $\alpha_{\text{in}} = 15$.

The distinction between fidelity and the conditional variance product based approaches is in some sense unsurprising. The T-V diagram and $\mathcal{M}$ are explicitly designed to be state independent, whilst fidelity is almost as explicitly state dependent. There are some consequences of this distinction regarding the experimetnal verification of the teleportation process. Characterization of T-V and $\mathcal{M}$ makes use of known test states to determine the transfer function of the teleporter. Provided the teleporter is not biased towards the test states, their particular form is irrelevant. On the other hand, characterization of fidelity uses a set of states representative of the assumed ensemble of unknown input states. In particular, for validity of the coherent state teleportation limits $\mathcal{F}_{\text{class}} = 1/2$ and $\mathcal{F}_{\text{no-cloning}} = 2/3$, the test states must have a sufficiently large range of coherent amplitudes to represent a broad distribution. The fidelity is then highly gain dependant, as we have seen from figs. 6 and 7. In this sense fidelity is a stronger test of quantum teleportation since it requires both high precision quantum and classical control. On the other hand, it becomes essential to accurately characterize the teleportation gain.

*E. Entanglement swapping*

To illustrate why non-unity gain teleportation is of interest we will consider the example of continuous variable entanglement swapping. Entanglement swapping utilizing a continuous variable teleportation protocol was first introduced by Polkinghorne and Ralph [16]. They showed that in a polarization teleportation protocol, effectively comprising of two quadrature teleporters, if the input is one of a pair of polarization entangled photons, then the output and the other polarization entangled photon can violate a Clauser-Horne-type inequality [32]. An interesting result of their work was that when weak continuous variable entanglement was used in the teleportation protocol, entanglement swapping could only be achieved for teleportation gain less than unity ($g^\pm < 1$). In this section we consider entanglement swapping of quadrature entanglement using a quadrature teleportation protocol. This type of entanglement swapping has been considered previously by Tan [33], by van Loock and Braunstein [34], and more recently by Zhang *et. al* [35].

Tan [33] considered entanglement swapping using a unity gain teleportation protocol ($g = 1$). He defined successful entanglement swapping to occur when the inequality $\langle (\delta \hat{X}_x^+ - \delta \hat{X}_y^+)^2 \rangle + \langle (\delta \hat{X}_x^- + \delta \hat{X}_y^-)^2 \rangle \geq 4$ is violated, where the sub-scripts $x$ and $y$ label the two sub-systems that the entanglement is being interrogated over. This is a sufficient criterion for the inseparability of sub-systems $x$ and $y$. Tan then showed that entanglement swapping occurs successfully if the squeezed beams used to generate the entanglement both used in, and input to, the teleportation protocol have at least 3 dB of squeezing. van Loock and Braunstein considered a more complex system, where the output entanglement from the entanglement swapping protocol is used to teleport a coherent state [34]. They allowed non-unity gain in the entanglement swapping teleporter, but not in the coherent state teleporter, and characterized the success of entanglement swapping by the fidelity of the coherent state teleportation as defined in eq. (5). Violation of the criterion $\mathcal{F} \leq 0.5$ is however, equivalent to violation of the criterion used by Tan, the significant difference between the two papers was the use of non-unity gain teleportation by van Loock and Braunstein. With this extra degree of freedom van Loock and Braunstein showed that entanglement swapping could be performed for any non-zero input squeezing. Zhang *et. al* [35] performed a similar analysis to that of van Loock and Braunstein but proposed an alternative experimental configuration.

The entanglement swapping protocols discussed above characterized success with sufficient but not necessary conditions for entanglement. Some situations in which the entanglement swapping was successful were therefore not identified. Here, we will consider entanglement swapping in more detail, using the inseparability criterion proposed by Duan *et al.* [36], [37] to characterize the success of the process. The inseparability criterion relies on the identification of separability with positivity of the P-function, and is a necessary and sufficient criterion for the presence of entanglement. For states with Gaussian noise distributions and symmetric correlations on the orthogonal quadratures, it can be related to measurable correlations [36]

$$\left\langle \left(k\delta\hat{X}_x^+ + \frac{\delta\hat{X}_y^+}{|k|}\right)^2 \right\rangle + \left\langle \left(k\delta\hat{X}_x^- - \frac{\delta\hat{X}_y^-}{|k|}\right)^2 \right\rangle < 2\left(k^2 + \frac{1}{k^2}\right) \quad (18)$$

where $k$ is an experimentally adjustable parameter. Note that if $k$ is restricted to unity this criterion becomes equivalent to the entanglement criterion used by Tan [33]. We define the *degree of inseparability* $\mathcal{I}$ as the product form of this inequality, so that our results here remain consistent with our experimental results presented later, and normalize so that the state is inseparable if $\mathcal{I} < 1$ [40]

$$\mathcal{I} = \frac{\left\langle \left(k_{\rm opt}\delta\hat{X}_x^+ + \frac{\delta\hat{X}_y^+}{|k_{\rm opt}|}\right)^2 \right\rangle^{\frac{1}{2}} \left\langle \left(k_{\rm opt}\delta\hat{X}_x^- - \frac{\delta\hat{X}_y^-}{|k_{\rm opt}|}\right)^2 \right\rangle^{\frac{1}{2}}}{k_{\rm opt}^2 + 1/k_{\rm opt}^2} \quad (19)$$

where $k_{\rm opt}$ is chosen to minimize $\mathcal{I}$. We use this measure to characterize the strength of quadrature entanglement between a pair of optical beams before one is sent through a teleporter $\mathcal{I}_{\rm initial}$, and afterwards $\mathcal{I}_{\rm final}$.

We utilize the teleportation protocol given in fig. 2, with the output defined by eqs. (2) and (3). The second quadrature entangled pair required for entanglement swapping, can be produced identically to the one used for teleportation, by combining two amplitude squeezed beams on a 50/50 beamsplitter. The entangled pair can then be described by the quadrature operators $\hat{Y}_x^\pm = (\hat{Y}_{\rm sqz,1}^\pm + \hat{Y}_{\rm sqz,2}^\mp)/\sqrt{2}$ and $\hat{Y}_y^\pm = (\hat{Y}_{\rm sqz,1}^\pm - \hat{Y}_{\rm sqz,2}^\mp)/\sqrt{2}$, where $\hat{Y}_{\rm sqz,1}^\pm$ and $\hat{Y}_{\rm sqz,2}^\pm$ are quadrature operators describing the amplitude squeezed beams used to generate the entanglement. We replace the operators describing the teleporter input state in eqs. (2) and (3) with those from entangled beam $x$; $\hat{X}_{\rm in}^\pm = \hat{Y}_x^\pm$. We then determine $\mathcal{I}_{\rm final}$ between the quadratures of the output from the teleportation protocol $\hat{X}_{\rm out}^\pm$ and those from the second entangled beam $\hat{Y}_y^\pm$. For simplicity, here we assume that the input entangled state is symmetric and pure so that $\Delta^2\hat{Y}_{\rm sqz} = \Delta^2\hat{Y}_{\rm sqz,1}^+ = \Delta^2\hat{Y}_{\rm sqz,2}^+ < 1$ and $\Delta^2\hat{Y}_{\rm anti} = \Delta^2\hat{Y}_{\rm sqz,1}^- = \Delta^2\hat{Y}_{\rm sqz,2}^- = 1/\Delta^2\hat{Y}_{\rm sqz} > 1$. $\mathcal{I}_{\rm final}$ as a function of the teleportation gain $g = g^+ = g^-$ is shown in fig. 8. Fig. 8 a) shows $\mathcal{I}_{\rm final}$ for $\mathcal{I}_{\rm initial} = 0.5$, ($\Delta^2\hat{Y}_{\rm sqz} = 0.5$) and for a range of teleportation efficacies (or in other words, a range of $\Delta^2\hat{X}_{\rm sqz}$). The degree of inseparability is, of course, degraded by the entanglement swapping procedure. We see, however, that the procedure is successful ($\mathcal{I}_{\rm final} < 1$) over a wide range of teleportation gains, and that, unlike the analysis of [16], it is always successful for $g = 1$. This difference is due to the more stringent nature of Clauser-Horne-type inequalities compared to tests of the presence of entanglement. Interestingly, optimal entanglement swapping ($\min_g\{\mathcal{I}_{\rm final}\}$) occurs at teleportation gain $g_{\rm opt}$ below unity in all cases except for the unphysical situations when $\Delta^2\hat{X}_{\rm sqz} \to 0$ or $\Delta^2\hat{Y}_{\rm sqz} \to 0$.

$$g_{\rm opt} = 1 - \frac{2(\Delta^2\hat{X}_{\rm sqz} + \Delta^2\hat{Y}_{\rm sqz})}{\Delta^2\hat{X}_{\rm sqz} + \Delta^2\hat{X}_{\rm sqz}^{-1}o + \Delta^2\hat{Y}_{\rm sqz} + \Delta^2\hat{Y}_{\rm sqz}^{-1}} \quad (20)$$

It can be shown that at this optimal point $k_{\rm opt} = 1$ so that the inseparability criterion of eq. (19) becomes equivalent to the criteria of Tan [33] and van Loock and Braunstein [34]. Unsurprisingly then, the optimum gain for entanglement swapping derived here is in agreement with that of van Loock and Braunstein [34]. For $g \neq g_{\rm opt}$ however, it is no longer true that $k_{\rm opt} = 1$, so that the gain bandwidth for successful entanglement swapping we present here is wider than what could be obtained using the schemes of Tan, or van Loock and Braunstein.

Fig. 8 b) shows the same result as fig. 8 a) but with stronger initial entanglement ($\mathcal{I}_{\rm initial} = 0.1$). Unsurprisingly, in this case the optimal final degree of inseparability is better than the previous example, and the optimal gain is closer to unity. Notice however that, for a given $\Delta^2\hat{X}_{\rm sqz}$, the gain bandwidth for successful entanglement swapping is the same in both Fig. 8 a) and b). Calculation of this gain bandwidth yields the same bandwidth as is given in eq. (18) for $\mathcal{M} < 1$. We see then that, indeed, the entanglement swapping gain bandwidth only depends on variances of the squeezed beams used in the teleportation protocol, and is independent of the strength of the input entanglement. The fact that the gain bandwidth for entanglement swapping corresponds exactly to that for $\mathcal{M}$ is a clear indication of the relevance of $\mathcal{M}$ as a measure for non-unity gain teleportation. In fact, if the input entanglement is perfect ($\Delta^2\hat{Y}_{\rm sqz} \to 0$), the degree of inseparability for the entanglement swapping protocol discussed here becomes equivalent to $\mathcal{M}$

$$\mathcal{I}_{\Delta^2\hat{Y}_{\rm sqz} \to 0} = \sqrt{\mathcal{M}} \quad (21)$$

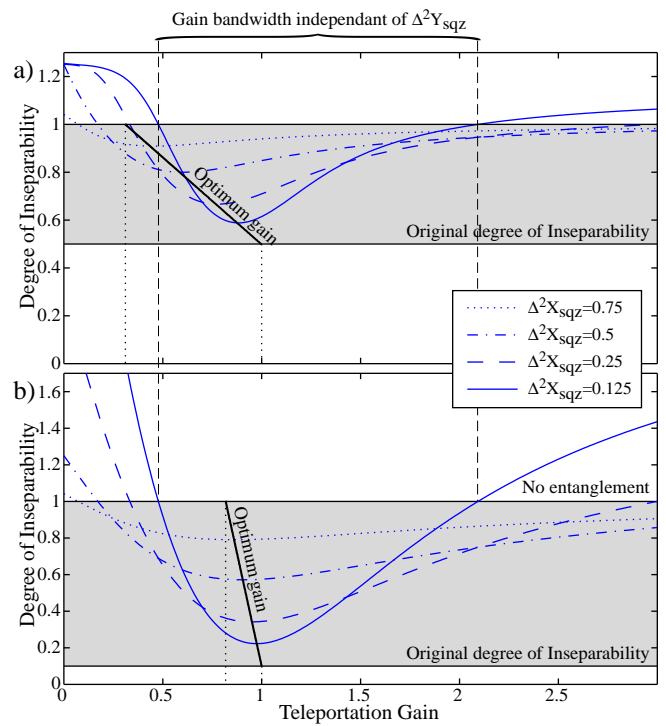

Fig. 8. Degree of entanglement between a pair of entangled beams after one is teleported as a function of the teleportation gain. a) $\Delta^2\hat{Y}_{\rm sqz} = 0.5$, b) $\Delta^2\hat{Y}_{\rm sqz} = 0.1$.

## IV. EXPERIMENT

This section describes our experimental teleportation protocol. In the following part we detail the process by which we generated quadrature entanglement, and discuss our characterization of this entanglement. In section IV-B we describe the teleportation protocol itself, and in section IV-C we analyze the protocol in terms of fidelity, signal transfer and conditional variance, and the gain normalized conditional variance. Finally, in section IV-D we give an example of the experimental loopholes associated with quantum teleportation.

### A. Generation of entanglement

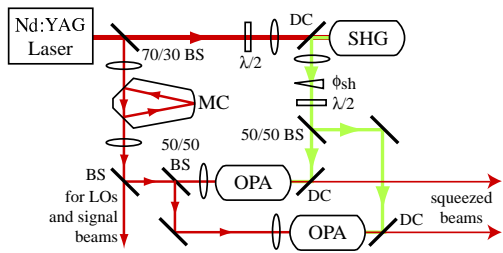

Fig. 9. Experimental apparatus used to generate two squeezed beams. BS: beam splitter, MC: mode cleaning resonator, DC: Dichroic, $\lambda/2$: half-wave plate, $\phi_{\rm sh}$: second harmonic phase shifter, LOs local oscillators.

The experimental apparatus for production of the pair of amplitude squeezed beams used in this work has been described in detail elsewhere [38]. A schematic diagram is given in fig. 9. The laser source was a 1.5 W monolithic non-planar ring Nd:YAG laser at 1064 nm. Roughly two thirds of it's output was frequency doubled with 50 % efficiency to produce 370 mW of 532 nm light. This was used to pump a pair of optical parametric amplifiers (OPAs). The remaining 1064 nm light was passed through a high finesse mode cleaning cavity to reduce its spectral noise and provided seeds for the two OPAs, as well as beams used to encode the input and output signals and a local oscillator for interrogation of the input and output states by Victor. Each OPA consisted of a hemi-lithic MgO:LiNbO$_3$ crystal and an output coupler. One end of the crystal had a 10 mm radius of curvature and was coated for high reflection at 1064 and 532 nm. The other end was flat and anti-reflection coated at both 1064 and 532 nm. The output couplers had 25 mm radii of curvature, were anti-reflection coated for 532 nm ($R_{532} \approx 7$ %), and had 96 % reflection of 1064 nm. In each OPA 23 mm separated the MgO:LiNbO$_3$ crystal and the output coupler, this created a cavity mode for the resonant 1064 nm light with a 27 $\mu$m waist at the center of the MgO:LiNbO$_3$ crystal. When locked to amplification, the 1064 nm output from each OPA exhibited phase squeezing, and when locked to de-amplification it exhibited amplitude squeezing. Our OPAs are quite susceptible to pick-up which couples noise directly into the phase quadrature, to avoid this problem we chose to lock to amplitude squeezing and observed 3.6 dB of squeezing at 8.4 MHz from each OPA via homodyne detection with roughly 84 % total efficiency. From this we infer 4.8 dB of squeezing directly after each OPA.

We produced quadrature entanglement by combining the two amplitude squeezed beams with a $\pi/2$ phase shift on a 50/50 beam splitter. We characterized the entanglement using two entanglement measures, the degree of inseparability given by eq. (19) (see section III-E), and the Einstein-Podolsky-Rosen (EPR) paradox criterion proposed by Reid and Drummond [39]. The EPR paradox criterion is given by $\mathcal{E} = \Delta^2 \hat{X}^+_{x|y} \Delta^2 \hat{X}^-_{x|y}$, where $\mathcal{E} < 1$ implies demonstration of the paradox. A detailed report on this characterization is given in [40]. The optimum observed value for the inseparability criterion was $\mathcal{I} = 0.44 \pm 0.02$ which is well below the limit for inseparability of unity. For our entanglement this value is equivalent to the average of the squeezed variances from the two OPAs. This corresponds to 3.6 dB of squeezing on each squeezed beam, in good agreement with our previously measured squeezing results. The optimum value of the EPR paradox criterion achieved was $\mathcal{E} = 0.58 \pm 0.02$. To characterize the optimum non-unity gain performance of our teleportation protocol possible with this entanglement resource we require, also, the mixedness $K$ of the entanglement. The inseparability criterion is independent of mixedness [40]. The EPR paradox criterion, however, has a strong dependence, and we found $K = \Delta^2 \hat{X}_{\rm sqz} \Delta^2 \hat{X}_{\rm anti} = 3.3$. The homodyne detector used to characterize our entanglement had 15 % loss. Inferring out this loss we find that $\mathcal{I}_{\rm tele} = 0.33$, and $K_{\rm tele} = 2.8$ for the entanglement directly upon entering the teleporter. We use these values in section IV-C to predict the theoretical optima of fidelity (fig. 13), T-V (fig. 15), and $\mathcal{M}$ (fig. 16) for our teleportation apparatus.

### B. Teleportation apparatus

Fig. 10 illustrates the optical (fig. 10 a)) and electrical (fig. 10 b)) configuration for our teleportation protocol. The apparatus consisted of three distinct parts: measurement (Alice), reconstruction (Bob), and generation and verification (Victor). At the generation stage Victor generated the input signal by independently amplitude and phase modulating an optical beam at 8.4 MHz. This produced a coherent state at 8.4 MHz with a coherent amplitude unknown to either Alice or Bob. He could then measure the Wigner function of this input state in a homodyne detector. Assuming that the input state is Gaussian however, Victor need only characterize the amplitude and phase quadratures to completely define the state. We make that assumption here, we lock Victor's homodyne to the amplitude quadrature using a Pound-Drever-Hall-type error signal [41], and to the phase quadrature by balancing the power to the two detectors incorporated in the homodyne detector (see fig. 10 b)). Victor then shared this fully characterized input state with Alice. Alice interfered it with one of the entangled beams with $\pi/2$ phase shift on a 50/50 beam splitter. The absolute intensities of the entangled beam and the input signal beam were arranged to be identical. Alice then detected the two beam splitter outputs with identical detectors, the detector darknoise was 10 dB below the quantum noise of the input beam. The sum (difference) of the two output photocurrents gave a measure of the amplitude (phase) quadrature of the signal, degraded by the amplitude (phase) quadrature of the entangled beam. Notice that it is this mixing of the fluctuations from the signal and the entanglement that allows quantum teleportation to be performed successfully. If Alice gains information about the quantum fluctuations of the input state through her measurements, then the Heisenberg uncertainty principle dictates that Bob's reconstruction must be degraded.

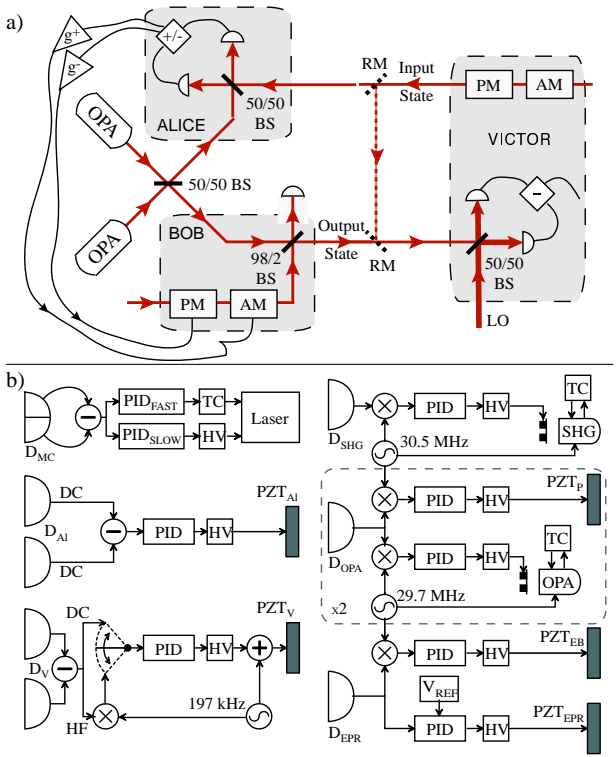

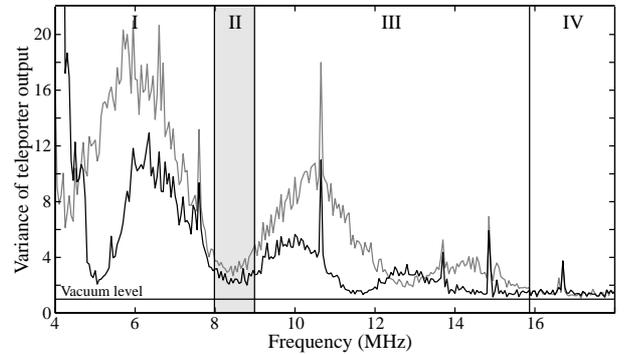

Fig. 10. a) Optical and b) electronic apparatus for the teleportation protocol. RM: removable mirror; HV: high voltage amp; PID: proportional, integral, differential locking servo; TC: temperature controller; $D_{SHG}$: SHG locking detector; $D_{OPA}$: OPA locking detector; $D_{EPR}$: entanglement and encoding beam locking detector; $D_{MC}$: mode cleaner locking detector; $D_{Al}$: Alice homodyne locking detector; $D_V$: Victor homodyne locking detector; PZT: piezoelectric crystal; $PZT_P$: OPA green phase PZT; $PZT_{EB}$: encoding beam PZT; $PZT_P$: OPA green phase PZT; $PZT_{EPR}$: entanglement locking PZT; $PZT_{Al}$: Alice's PZT; $PZT_V$: Victor's PZT.

Alice sent her two photocurrents to Bob. Bob could then, if he wanted, simply encode those photocurrents using an amplitude and a phase modulator on his entangled beam. Typically however, the loss introduced by modulators is non-trivial, and would degrade Bob's reconstruction of the input state. Instead we applied the photocurrents to a bright coherent optical beam, and combined this encoding beam and Bob's entangled beam with controlled phase on a 98/2 beam splitter. One beam splitter output was Bob's reconstructed output state. This scheme avoids the loss introduced by the modulators, and the only loss introduced to the entanglement by Bob is then the 2 % due to the beam splitter ratio. The attenuation of the signals modulated onto the encoding beam caused by the 98/2 beam splitter was counter-acted by Bob simply increasing the gain of his encoding by a factor of 50.

Bob then provided Victor with his reconstructed state. Victor used a removable mirror to switch his homodyne detector input to the output state. By making amplitude and phase quadrature measurements, he was then able to fully characterize the output state, and judge how well the teleportation had been performed. Normally inefficiencies in the teleportation protocol would be expected to degrade Victor's judgement of how well the process has been performed. There is one notable exception to this rule however, that of loss in Victor's homodyne. This loss appears erroneously to Victor to enhance the process, and must be accounted for. The combined loss from Victor's homodyne mode-matching and detector photodiodes was characterized and found to be 15 %±2 %. This loss was inferred out of the final results obtained by Victor.

Fig. 11. Spectral variance of the amplitude $\Delta^2 \hat{X}^+_{out}$ (dark trace) and the phase $\Delta^2 \hat{X}^-_{out}$ (light trace) quadratures of the teleported signal, normalized to the shot noise of Victor's homodyne detector. The teleportation gain and feed-forward phase vary over the spectra dependent on the response function of our detectors and feed-forward loops. I: $g > 1$, II: $g \approx 1$, III: $g < 1$, IV: $g \approx 0$.

To characterize the behavior of our detectors and feed-forward loop we analyzed the amplitude and phase quadrature frequency spectra of the output state received by Victor over a wide frequency range, with $g_{8.4\ MHz} \approx 1$ and no input coherent amplitude. These frequency spectra are presented in fig. (11). The transfer functions of the detectors and feed-forward loop, and the frequency dependence of our entanglement all have a bearing on the output spectra. A definite phase relationship must be maintained between the fed-forward signal and the fluctuations of the second entangled beam to get maximum cancellation of the fluctuations from the entanglement. The time delay in our electronics and detectors caused cycling of the relative phase, this results in cycling between maximum and minimum cancellation of the entangled beam fluctuations. We see this in fig. (11) as a sinusoidal modulation of both the amplitude and phase spectra with 3-4 MHz period. We arranged the feed-forward phase so that a maximum cancellation point occurred at 8.4 MHz for both the amplitude and phase quadratures. The amplitude of the sinusoidal modulation depends on both the amplitude response of our electronics and the size of the quadrature fluctuations of our entangled beams. Over the frequency range of the measurements presented in fig. (11) both the response of our electronics, and the amplitude and phase quadrature variances of our entanglement ($\Delta^2 \hat{X}^\pm_x$ and $\Delta^2 \hat{X}^\pm_y$) decrease with increasing frequency. At low frequencies (region I) we then have $g > 1$ and this, coupled with the large variances of our entangled beams, results in the sinusoid having a large amplitude. Around 8.4 MHz (region II) $g \approx 1$ and the feed-forward phase is optimized, so that $\Delta^2 \hat{X}^\pm_{out}$ are minimized, this is the region in which we performed our teleportation protocol. At higher frequencies (region III) $g < 1$ so that the feed-forward has a smaller effect on the output variances, and in region IV there is effectively no feed-forward ($g \approx 0$) and the output variances is simply the variance of Bob's entangled beam.

## C. Teleportation results

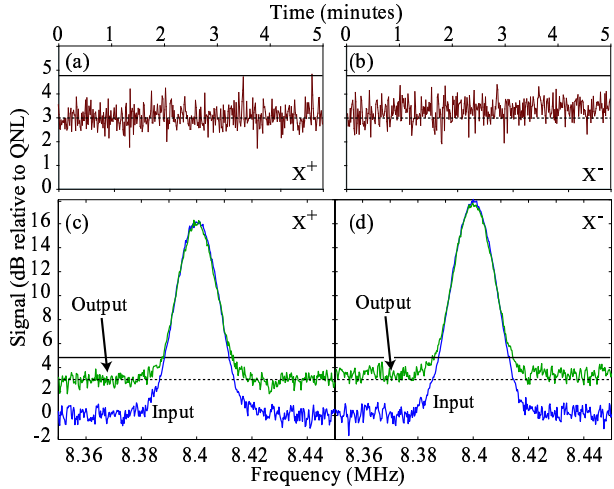

Fig. 12. The input and output states of the teleporter, as measured by Victor. (a) and (b) show the amplitude and phase noise of the output state at 8.4 MHz. (c) and (d) show the input and output of the teleporter, when probed with a signal at 8.4 MHz. In all cases, the dotted line is the no-cloning limit, while the solid line is the classical limit. All data has been corrected to account for the detection losses of Victor. Resolution Bandwidth=10 kHz, Video Bandwidth=30 Hz.

A sample of the data obtained from our teleporter is shown in Fig. 12. Parts (a) and (b) show the amplitude and phase quadrature noise of the output state at 8.4 MHz measured by Victor, as a function of time. The complete system maintained lock for long periods. Given the assumption that all noise sources introduced during the teleportation process, including the entanglement and the input state, were Gaussian, fig. 12 (c) and (d) contain sufficient information to fully characterize the teleportation run. Every teleportation run consisted of four spectra, such as these, as well as a quantum noise calibration (not shown). Also drawn in each part of fig. 12 are lines corresponding to the classical limit (solid line, +4.8 dB) and the no-cloning limit (dashed line, +3 dB). The data in (c) and (d) show Victor's measurement of the amplitude and phase quadratures over a 100 kHz bandwidth centered around 8.4 MHz. Over this range the noise floor of the system was constant, which could be easily verified by switching the coherent amplitude of the input state off and on. We obtained $\Delta^2 \hat{X}_{in}^{\pm}(8.4 \text{ MHz})$ and $\Delta^2 \hat{X}_{out}^{\pm}(8.4 \text{ MHz})$ respectively, from the average of $\Delta^2 \hat{X}_{in}^{\pm}(\omega)$ and $\Delta^2 \hat{X}_{out}^{\pm}(\omega)$ at nearby frequencies. Henceforth, if the $\omega$ in expressions such as these is neglected it implies that the measurement is at 8.4 MHz, (for example $\Delta^2 \hat{X}_{in}^{\pm} = \Delta^2 \hat{X}_{in}^{\pm}(8.4 \text{ MHz})$).

$$\Delta^2 \hat{X}_{in}^{\pm} = \frac{\int_{8.35 \text{ MHz}}^{8.37 \text{ MHz}} \Delta^2 \hat{X}_{in}^{\pm}(\omega) d\omega + \int_{8.43 \text{ MHz}}^{8.45 \text{ MHz}} \Delta^2 \hat{X}_{in}^{\pm}(\omega) d\omega}{0.04 \text{ MHz}}$$

$$\Delta^2 \hat{X}_{out}^{\pm} = \frac{\int_{8.35 \text{ MHz}}^{8.37 \text{ MHz}} \Delta^2 \hat{X}_{out}^{\pm}(\omega) d\omega + \int_{8.43 \text{ MHz}}^{8.45 \text{ MHz}} \Delta^2 \hat{X}_{out}^{\pm}(\omega) d\omega}{0.04 \text{ MHz}}$$

Using $\Delta^2 \hat{X}_{in}^{\pm}$ and $\Delta^2 \hat{X}_{out}^{\pm}$ it was then possible to extract the input and output coherent amplitudes $\alpha^{\pm}$ from the input $\Delta^2 \hat{M}_{in}^{\pm}$ and output $\Delta^2 \hat{M}_{out}^{\pm}$ variance measurements at 8.4 MHz.

$$\alpha_{in}^{\pm} = \frac{1}{2}\sqrt{\Delta^2 \hat{M}_{in}^{\pm} - \Delta^2 \hat{X}_{in}^{\pm}} \quad (22)$$

$$\alpha_{out}^{\pm} = \frac{1}{2}\sqrt{\Delta^2 \hat{M}_{out}^{\pm} - \Delta^2 \hat{X}_{out}^{\pm}} \quad (23)$$

The amplitude and phase quadrature gains of the teleportation could then be directly calculated ($g^{\pm} = \alpha_{out}^{\pm}/\alpha_{in}^{\pm}$), and all of the measures of teleportation discussed in section III could be obtained.

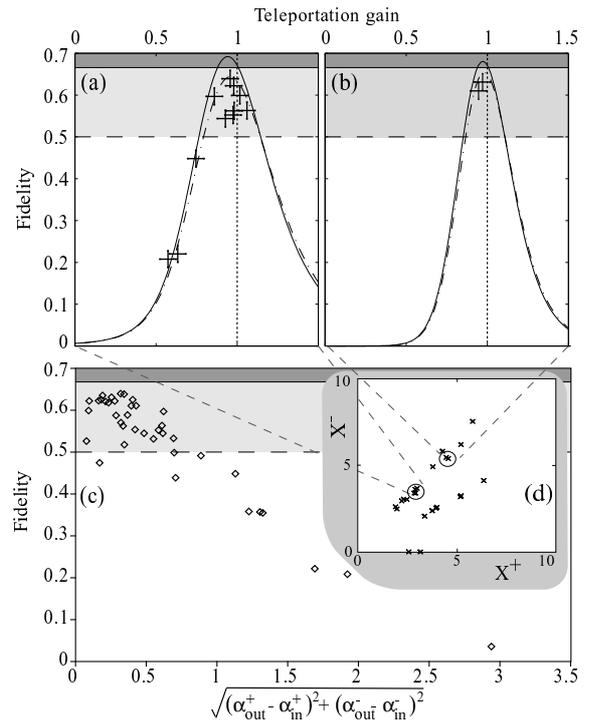

Fig. 13. Measured fidelity plotted; versus teleportation gain, $g$, in (a) and (b); versus coherent amplitude separation between input and output states in (c); and on phase space in (d). In (a) the input signal size was $(\alpha^+, \alpha^-) \approx (2.9, 3.5)$ and in (b) $(\alpha^+, \alpha^-) \approx (4.5, 5.4)$. $g$ was calculated as the ratio of the input and output coherent amplitudes. The dashed (solid) lines show the classical (no-cloning) limits of teleportation at unity gain. The solid curves are calculated optima based on the characterization of our entanglement in section IV-A, and the efficiency of the protocol. The dot-dashed curves include the experimental asymmetric gains: for (a) $g^- = 0.84 g^+$ and for (b) $g^- = 0.92 g^+$.

*1) Fidelity analysis:* The fidelity results obtained from all of our teleportation runs are displayed in fig. 13. As discussed in section III-A, an important parameter of any study of fidelity is the area of phase space from which the inputs states are produced. Fig. 13(d) shows the area of phase space that our experiment has probed. All points shown here satisfied $\mathcal{F} > 0.5$. This area was limited in radius by the increasing sensitivity of fidelity to gain as $\alpha_{in}$ increases (see fig. 3), and small $\alpha_{in}^{\pm}$ were avoided so that the teleportation gain of both quadratures could be accurately verified. A summary of all our fidelity results is shown in fig. 13 (c) as a function of deviation from unity gain. Fig. 13 (a) and (b) each display a subset of our fidelity results for input states with particular coherent amplitudes. The solid curves show the best possible performance of our system, based on our entanglement, detection efficiency, dark noise, and assuming equal gain on each quadrature. In both plots the highest fidelity occurs for gain less than unity. The increased fidelity is less obvious in (b) where the signal is approximately twice as large as that in (a). In fact, in the limit of a vacuum input state, the fidelity criterion will be satisfied perfectly by a classical teleporter (i.e. one with the entangled state replaced by two coherent states) with zero gain. These results demonstrate the necessity of obtaining and verifying the correct gain settings for any implementation of continuous variable teleportation. Obtaining the correct gain setting is actually one of the more troublesome experimental details. To illustrate this point, we have plotted the dashed curves on (a) and (b) for a teleporter with asymmetric quadrature gains. Such asymmetry was not unusual in our system, and explains the variability of the results shown in fig. 13 (a).

In our experiment the amplitude and phase quadrature teleportation gains were adjusted to the desired level by encoding a large coherent modulation on the input state ($\alpha_{\text{in}}^{\pm} \gg \sqrt{\Delta^2 \hat{X}_{\text{in}}^{\pm}}$ and $\alpha_{\text{out}}^{\pm} \gg \sqrt{\Delta^2 \hat{X}_{\text{out}}^{\pm}}$). The modulation transferred to the output state was then measured, since the noise on the input and output states was negligible compared to the signal the gain could be obtained directly $g^{\pm} \approx \sqrt{\Delta^2 \hat{M}_{\text{out}}^{\pm}/\Delta^2 \hat{M}_{\text{in}}^{\pm}}$, and then optimized to the desired level. In some experiments characterization such as this were used to yield the final teleportation gain used to calculate fidelity. In the work of Zhang et al. [10] they adjust the gain to unity, and then assume that it remains at unity throughout the teleportation run. In a perfect experimental situation this procedure will work perfectly well. In our experiment however, we found that no matter how well the gain was set initially, the teleportation gain would drift slightly during the course of an experimental run. Since the fidelity is extremely sensitive to gain (see fig. 3) even very small drifts in the gain can lead to significant degradation. We therefore believe that it is important to experimentally verify the gain of teleportation during each teleportation run. Fig. 14 illustrates this point, displaying a histogram of the complete set of our teleportation fidelity results calculated firstly using the teleportation gain measured during each run (fig. 14 a)), and secondly by assuming that the gain was adjust correctly beforehand to unity (fig. 14 b)). We see quite significant differences in the two distributions. The maximum fidelity we observed with experimentally verified gain was $\mathcal{F} = 0.64 \pm 0.02$ with $g = 1.02 \pm 0.05$, without experimentally verifying the gain however, we thrice observed $\mathcal{F} > 2/3$.

*2) Signal transfer and conditional variance analysis:* As discussed in section III, there are merits to analyzing teleportation in a state independent manner. Here, we present our teleportation results analyzed in terms of signal transfer and conditional variances between the input and output states. We display the results on a T-V diagram as described in section III-B, and also present them in terms of the gain normalized conditional variance measure $\mathcal{M}$ introduced in section III-C.

Our T-V results are shown in fig. 15. The classical limit curve shows the ideal achievable result as a function of gain if the entanglement was replaced with two coherent states, and the efficiency of the protocol was unity. The unity gain curve shows the locus of points obtained at unity teleportation gain with increasing entanglement. Finally, a theoretical optimum (as a function of gain) is shown for our experimental parameters. By varying our experimental conditions, particularly the gain, we have mapped out some portion of the T-V graph. Perhaps the most striking feature of these results are the points with $T_q > 1$, the best of which has $T_q = 1.06 \pm 0.03$. Since only one party may have $T_q > 1$, this shows that Bob has maximal information about the input signal and we have broken the information cloning limit. The lowest observed conditional variance product was $V_q = 0.96 \pm 0.10$. This point also had $T_q = 1.04 \pm 0.03$. This is the first observation of both $T_q > 1$ and $V_q < 1$, as well as the first simultaneous observation of both. With unity gain simultaneously observing both $T_q > 1$ and $V_q < 1$ would imply breaking of the no-cloning limit for teleportation (i.e $\mathcal{F} > 2/3$). This particular point, however, had a fidelity of only $0.63 \pm 0.03$. The main reason for this low fidelity is asymmetric gain, the amplitude gain was $g^+ = 0.92 \pm 0.08$ while the phase gain was $g^- = 1.12 \pm 0.08$. Such gain errors have a dramatic impact on the measured fidelity because the output state then has a different classical amplitude ($\alpha^\pm$) to the input. On the other hand, they have only a minor effect on $T_q$ and $V_q$.

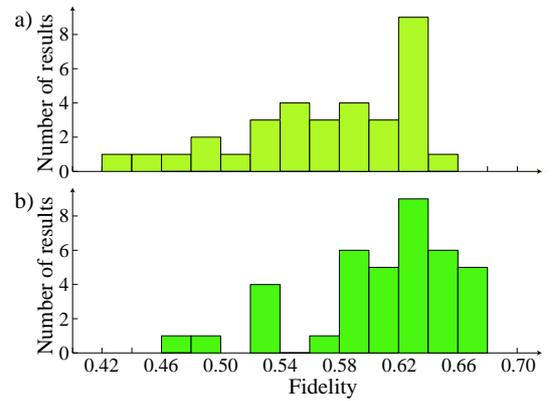

Fig. 14. Histogram of the fidelity obtained from our teleportation experiment comparing fidelities obtained a) when the gain is experimentally verified, and b) when unity teleportation gain is assumed.

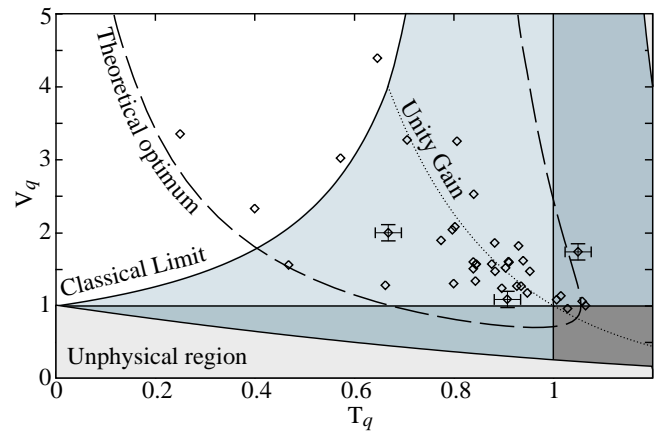

Fig. 15. T-V graph of the experimental results. The dashed theoretical optimum was calculated based on the characterization of entanglement in section IV-A and experimental losses. Representative error bars are shown for some points.

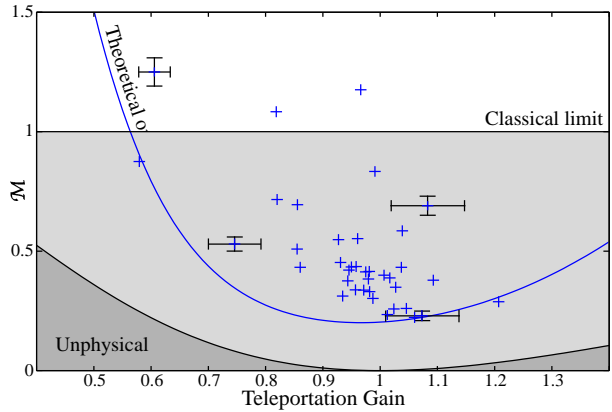

Fig. 16. Plot of our experimental $\mathcal{M}$ results as a function of teleportation gain. Representative error bars are shown for some points. The theoretical optimum was calculated based on the characterization of entanglement in section IV-A and experimental losses; and the unphysical region is arrived at from eq. (16).

Fig. 16 shows $\mathcal{M}$ as a function of teleportation gain for our experimental results. As discussed in sections III-C and III-E, $\mathcal{M}$ has real physical significance. An observation of $\mathcal{M} < 1$ implies not only that entanglement must have been used in the teleportation protocol, but also that the protocol could be used to perform entanglement swapping. We have experimentally demonstrated $\mathcal{M} < 1$ for a wide range of gains from $g_{\min} = 0.58$ to $g_{\max} = 1.21$, and achieved an optimum of $\mathcal{M} = 0.22 \pm 0.02$ at $g = 1.06 \pm 0.07$.

### D. An experimental loophole: single quadrature modulation

A number of loopholes exist in experimental demonstrations of quantum teleportation, and to some degree, any demonstration relies on the integrity of Alice, Bob and Victor. We consider the example of single quadrature modulation here, as shown in fig. 17. Here Victor has encoded no signal on the phase quadrature of the input state ($\alpha_{\text{in}}^- = 0$). Somehow, Bob has discovered this is the case. Bob can then reduce his phase quadrature teleportation gain below unity to minimize the phase reconstruction noise ($\min_g\{V_q\}$) with no fidelity cost. In the case of a classical teleportation protocol Bob could simply turn his phase quadrature feed-forward off, and it would appear to Victor that the phase quadrature had been reconstructed perfectly. In this example, however, the teleportation protocol does utilize entanglement, and the optimum strategy for Bob is to leave the phase quadrature feed-forward on, just with reduced gain. Fig. 17 c) shows the strong asymmetry then created between the reconstruction of the amplitude and phase quadratures. Victor, when analyzing the fidelity finds an artificially enhanced value of $\mathcal{F}' = 0.70$. We see that it is essential for Alice and Bob to obtain no information about the orientation of the input coherent amplitude. In our experiment this was achieved by spatially isolating Victor's station from those of Alice and Bob, and by varying the amplitude and phase quadrature coherent amplitudes $\alpha_{\text{in}}^\pm$ independently.

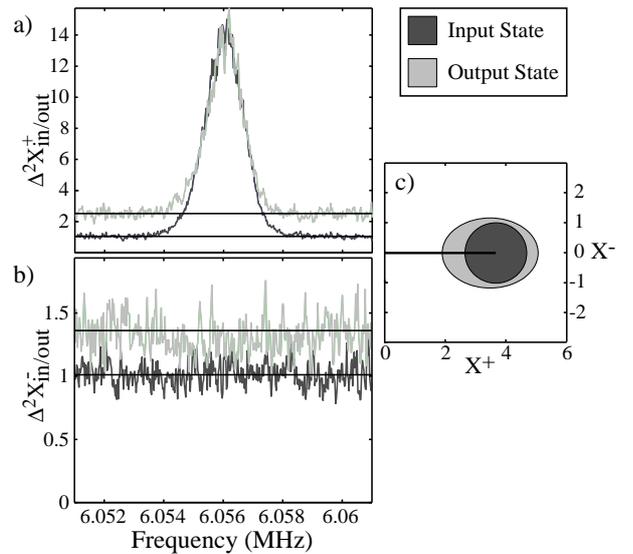

Fig. 17. Experimental demonstration of teleportation loophole when no phase signal is applied. a) and b) show the amplitude and phase noise of the output state at 8.4 MHz; and c) shows the standard deviation contours of the Wigner functions of the input and output states inferred from the measurements in a) and b).

### V. CONCLUSION

We have performed stably locked quantum teleportation of the amplitude and phase quadratures of an optical field. We characterized the teleportation using fidelity, a T-V diagram, and a measure derived in this paper - the gain normalized conditional variance product. The optimum directly observed fidelity was $\mathcal{F} = 0.64 \pm 0.02$. This was limited, not by the strength of our entanglement resource, or by the efficiency of the teleportation protocol, but rather by the stability with which control of the teleportation gain was possible. This can be seen from our T-V analysis, which was much less sensitive to gain. The maximum two quadrature signal transfer for our apparatus was $T_q = 1.06 \pm 0.03$; and we observed a conditional variance product of $V_q = 0.96 \pm 0.10$ and signal transfer of $T_q = 1.04 \pm 0.03$, simultaneously. These results are the first observation of $T_q > 1$ and $V_q < 1$, as well as the first simultaneous observation of both criteria. At unity gain simultaneously observing $T_q > 1$ and $V_q < 1$ ensures violation of the no-cloning limit for teleportation. The asymmetry in our gain, however, prevented a direct measurement of $\mathcal{F} > 2/3$. Based on the work of Grosshans and Grangier [1], and Ralph and Lam [2], we have derived a new measure for non-unity gain teleportation, the gain normalized conditional variance product $\mathcal{M}$. We analyze our teleportation results using this measure and demonstrate teleportation for gains from $g_{\min} = 0.58$ to $g_{\max} = 1.21$, achieving an optimum of $\mathcal{M} = 0.22 \pm 0.02 < 1$. We consider entanglement swapping characterized by a necessary and sufficient condition for entanglement, and demonstrate that the range of gains for which it is successful is dictated by $\mathcal{M}$, with an optimum, always, at $g < 1$.

## VI. Acknowledgements

We thank Hans-A. Bachor and Aska Dolinska for useful discussion, the Australian Research Council for financial support, and the Alexander von Humboldt foundation for support of R. Schnabel. This work is a part of EU QIPC Project, No. IST-1999-13071 (QUICOV).

**Warwick P. Bowen** Warwick Bowen graduated from the University of Otago, New Zealand, in 1998. He is currently undertaking his PhD at the Australian National University, a part of which was to design and implement the quantum teleportation experiment reported here. In 2002 he was a vistor in Physics in the Kimble laboratories at the California Institute of Technology. His research interests include the interface between quantum and atom optics.

**Nicolas Treps** Nicolas Treps undertook his phD in Paris, at the Laboratoire Kastler Brossel, on the quantum study of optical images. In 2002 he held a post-doctoral position at the Australian National University where he reseached continuous variable quantum information protocols and further quantum imaging experiments. He is now a lecturer at the University Pierre et Marie, doing his research within the Laboratoire Kastler Brossel.

**Ben C. Buchler** Ben Buchler obtained his BSc and PhD at the Australian National University, where his research was centred around quantum non-demolition measurements. He is currently working on near-field imaging of photonic crystals at ETH Zurich.

**Roman Schnabel** Roman Schnabel is currently scientific staff at the Max-Planck-Institute for Gravitational Physics (Albert-Einstein-Institute) in Hannover, Germany. His research interests cover experiments with squeezed light and advanced techniques for third generation gravitational wave detectors. In 2000, he received the Feodor Lynen fellowship of the Alexander von Humboldt Foundation.

**Timothy C. Ralph** Tim Ralph is a graduate of Macquarie University, Sydney, Australia and obtained his PhD from the Australian National University. He currently holds a Queen Elizabeth II Fellowship at the University of Queensland and is a Research Manager in the Centre for Quantum Computer Technolgy.

**Thomas Symul** Thomas Symul graduated from the Ecole Nationale Superieure des Telecommunications of Paris in 1998, and received is PhD degree in Physics from the University of Paris VI in 2001. He currently holds a post-doctoral position in the Quantum Optics group, Department of Physics, Australian National University.

**Ping Koy Lam** Ping Koy Lam obtained his BSc from the University of Auckland. He worked as an engineer for Sony and Hewlett Packard before his M. Sc. and PhD. studies at the Australian National University. He was awarded the 1999 Bragg Medal in Physics by the Australian Institute of Physics. He is presently a physics reader and the group leader of quantum optics research at the Australian National University. His research interests include nonlinear and quantum optics.